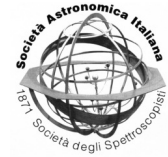

# The link between short Gamma–ray bursts and Gravitational Waves: perspectives for the THESEUS mission

P. D'Avanzo[1]

Istituto Nazionale di Astrofisica – Osservatorio Astronomico di Brera, Via Bianchi 46, I-23807 Merate (LC), Italy e-mail: `paolo.davanzo@brera.inaf.it`

**Abstract.** The knowledge of the class of short Gamma-Ray Bursts (GRBs), characterised by a duration of the gamma-ray emission ≤ 2 s, experienced an impressive boost in the last decade. In particular, the discovery of short GRB afterglows in 2005 with Swift and HETE-II provided the first insight into their energy scale, environments and host galaxies. The lack of detection of associated supernovae proved that they are not related to the death of massive stars. The increasing evidence for compact object binary progenitors makes short GRBs one of the most promising sources of gravitational waves for the forthcoming Advanced LIGO/Virgo science runs. To this end, the spectacular detection of the first electromagnetic counterpart of the gravitational wave event GW 170817 originated by the coalescence of a double neutron star (NS) system, represents a first hystorical milestone. The (weak) short GRB 170817A associated to this GW event provided the long-sought evidence that at least a fraction of short GRBs are originated by NS-NS merging and suggested the intriguing possibility that relativistic jets can be launched in the process of a NS-NS merger. The THESEUS mission, thanks to the diversity of intstrumentation, fast pointing and flexible schedule will represent a key facility in the multi-messenger astronomy era.

**Key words.**

## 1. Introduction

Gamma-ray bursts (GRBs) are rapid, powerful flashes of radiation peaking in the gamma-ray band, occurring at an average rate of one event per day over the whole sky at cosmological distances. The gamma-ray prompt emission is followed by a broadband (X-rays to radio ranges) fading afterglow emission that can be observed up to weeks and months after the onset of the event. Two classes of GRBs (at least), short and long, have been identified. Short GRBs (SGRBs) are identified as those with duration less than about two seconds and with harder spectra with respect to long bursts (Kouveliotou et al. 1993). These two classes of long ($T_{90} > 2$ s) and short GRBs ($T_{90} \leq 2$ s), show substantial evidences for different origins. Long GRBs, or at least a significant fraction of the nearby events (with redshift $z \leq 1$) for which it has been possible to search for the presence of a supernova (SN), are associated with the core-collapse explosions of massive stars (see Hjorth & Bloom 2012, for a recent review). Concernig short GRBs, current models suggest that they are associated with the merging of compact objects in binary systems, such as a double neutron star (NS), or a NS and a black hole (BH) system (Eichler et al. 1989; Narayan et al. 1992; Nakar 2007). These systems can originate from the evolution of massive stars in a primordial binary (Narayan et al. 1992) or by dynamical interactions in globular clusters during their core collapse (Grindlay et al. 2006; Salvaterra et al. 2008).

Short and long GRBs are not distinguished only by their duration and spectral hardness. Short GRBs are found to be typically less energetic (their isotropic equivalent energy, $E_{iso}$, is of the order of $10^{49} - 10^{51}$ erg) than long GRBs and to occur at a lower redshift (Nakar 2007; Berger 2011; Fong et al. 2013). Their afterglows tend to be significantly fainter on average than those of long GRBs (Kann et al. 2011; Nicuesa Guelbenzu et al. 2012; Margutti et al. 2013). Concerning the host galaxies, short GRBs occur in both early and late type galaxies with low star formation rate and are associated with an old stellar population (Berger 2009; Leibler & Berger 2010; Fong et al. 2013). A different origin for short GRBs with respect to the long GRB class is also supported by the lack of detection of the underlying supernova in the light curves of their optical afterglows down to very stringent magnitude limits (Hjorth et al. 2005a; Hjorth et al. 2005b; Fox et al. 2005; Covino et al. 2006; Kann et al. 2011; D'Avanzo et al. 2009) and by their inconsistency with the correlation, valid for long GRBs, between the rest frame spectral peak energy and $E_{iso}$ ($E_{peak} - E_{iso}$ correlation; Amati et al. 2002). On the other hand, Ghirlanda et al. (2009) showed that short GRBs are consistent with the same $E_{peak} - L_{iso}$ correlation (where $L_{iso}$ is the prompt emission isotropic peak luminosity) defined by long GRBs (Yonetoku et al. 2004). The distributions of the intrinsic X-ray absorbing column densities of long and short GRBs do not show significant differences when compared in the same redshift range ($z \lesssim 1$; Kopac et al. 2012; Margutti et al. 2013; D'Avanzo et al. 2014). Although all indirect, the observational evidences listed above are in good agreement with the expectations of the compact merger scenario (see Berger 2014 and D'Avanzo 2015 for recent reviews). Such a scenario has been spectacularly con-firmed on Aug 17 2017, when the first NS-NS gravitational wave event ever detected by aLIGO/Virgo (GW 170817) and associated to the weak short GRB 170817A detected by the *Fermi* and *INTEGRAL* satellites (Abbott et al. 2017a; Goldstein et al. 2017; Savchenko et al. 2017).

## 2. GW 170817/GRB 170817A: the dawn of the multi–messenger astronomy era

A gravitational wave (GW) event originated by the merger of a binary neutron star (BNS) system was detected for the first time by aLIGO/Virgo (GW 170817; Abbott et al. 2017a), and it was found to be associated to the weak short GRB 170817A detected by the *Fermi* and *INTEGRAL* satellites (Goldstein et al. 2017; Savchenko et al. 2017), marking the dawn of multi–messenger astronomy (Abbott et al. 2017b). The proximity of the event ($\sim$ 41 Mpc; Cantiello et al., 2018; Hjorth et al. 2017) and the relative accuracy of the localization ($\sim$ 30 deg$^2$, thanks to the joint LIGO and Virgo operation) led to a rapid ($\Delta t < 11$ hr) identification of a relatively bright optical electromagnetic counterpart (EM), named AT2017gfo, in the galaxy NGC 4993 (Arcavi et al. 2017; Coulter et al. 2017; Lipunov et al. 2017; Melandri et al. 2017; Soares-Santos et al. 2017; Tanvir et al. 2017; Valenti et al. 2017). The analysis and modelling of the spectral characteristics of this source, together with their evolution with time, resulted in a good match with the expectations for a "kilonova" (i.e. the emission due to radioactive decay of heavy nuclei produced through rapid neutron capture; Li & Paczyński 1998), providing the first compelling observational evidence for the existence of such elusive transient sources[1] (Cowperthwaite et al. 2017; Drout et al. 2017; Evans et al. 2017; Kasliwal et al. 2017; Nicholl et al. 2017; Pian et al. 2017; Smartt et al. 2017;

---

[1] Before GW 170817/GRB 170817A kilonova signatures have been tentatively identified in a few short GRBs light curves (Tanvir et al. 2013; Berger et al. 2013; Jin et al. 2016; Yang et al. 2015; Jin et al. 2015).

Villar et al. 2017). While the kilonova associated with GW 170817 has been widely studied and its main properties relatively well determined, the properties of the short GRB associated to GW 170817 appear puzzling in the context of observations collected over the past decades. The prompt $\gamma$–ray luminosity was significantly fainter (by a factor about 2500) than typical short GRBs (see, e.g., D'Avanzo et al. 2014). A faint afterglow was detected in the X–ray and radio bands only at relatively late-times (~9 and 16 d after the GW/GRB trigger, respectively; Alexander et al. 2017; Haggard et al. 2017; Hallinan et al. 2017; Margutti et al. 2017; Troja et al. 2017a). The extremely low $\gamma$-ray luminosity of GRB 170817A, together with the late-time rise of the afterglow has been interpreted as due to the debeamed radiation of a jet (uniform or structured[2] observed off–beam (i.e. viewing angle $\theta_{view} > \theta_{jet}$) or to a jet–less/chocked jet isotropic fireball with some stratification in its radial velocity structure (Pian et al. 2017; Lazzati et al. 2017b; Salafia et al. 2017; D'Avanzo et al. 2018; Lyman et al. 2018; Margutti et al. 2018; Mooley et al. 2018; Ruan et al. 2018; Salafia et al. 2018; Troja et al. 2018). Continued monitoring of the GRB 170817A afterglow carried out in radio, optical and X-ray bands carried out until ~ 160 d after the GW/GRB trigger provide a good agreement with the expectations for the structured jet and the isotropic fireball with a velocity profile (D'Avanzo et al. 2018; Lyman et al. 2018; Margutti et al. 2018; Mooley et al. 2018; Ruan et al. 2018). Only future observations can provide constraints to discriminate between these two scenarios. On the other hand, it is clear already at the present stage that the detection of GW170817/GRB 170817A, besides representing an historical result, clearly highlights the enormous scientific potential represented by the short GRB - GW connection that can be exploited in the next, forthcoming, LIGO/Virgo observing runs.

---

[2] In the structured jet scenario, the jet has a fast and energetic inner core surrounded by a slower, less energetic layer/sheath/cocoon (Lipunov et al. 2001; Rossi et al. 2002; Salafia et al. 2015, Kathirgamaraju et al. 2017, Lazzati e tal. 2017a; Gottlieb et al. 2017)

## 3. Perspectives for THESEUS

The case of GW170817/GRB 170817A highlights the importance of all-sky/wide-field high energy space-based facilities operating at the same time with gravitational waves interferometers. On one hand, the detection of an electromagnetic counterpart at high-energy can significantly reduce the uncerainties in source position in the sky. On the other hand, a (nearly) simultaneous detection of both the gravitational and the electromagnetic signal is fundamental to ensure a secure association of the two phenomena. To this end, the two high–energy instrument of THESEUS, namely the Soft X–ray Imager (SXI) and the X-Gamma rays Imaging Spectrometer (XGIS) can ensure a large field of fiew (1 and 1.5 sr, respectively), a broad-band coverage in terms of energy band (from 0.3 keV up to 20 MeV) and an accuracy in source location of a few arcmin (Amati et al. 2017). Furthermore, the THESEUS InfraRed Telescope (IRT), besides providing an accuracy in source location of less than 1", can be a key instrument to carry out narrow-field targeted search of GW electromagnetic counterparts (see, e.g, Evans et al. 2017) or to carry out time-resolved follow-up. Indeed, given the expected performances in terms of limiting magnitude (Amati et al. 2017), with IRT it woud have been possible to monitor the near-infrared light curve of the kilonova associated to GW 170817 (located at a distance of 41 Mpc) for ~ 15-20 days after the GW trigger (see, e.g., Tanvir et al. 2017). The same kilonova, could have been followed by IRT for ~ 7-10 days if located at a further distance (up to 100 Mpc). A multi-wavelenght monitoring carried out in the ultraviolet, optical and near-infrared is fundamental to discriminate between different kilonova models, enabling a determination of parameters like mass, velocity and composition of the ejecta (Pian et al. 2017, Evans et al. 2017, Tanvir et al. 2017, Villar et al. 2017).

## 4. Conclusions

The detection of the first electromagnetic counterpart of a gravitational wave event marked the dawn of the multi–messenger astronomy era. The single (by now) case of GW170817/GRB 170817A, besides representing an historical result, clearly demonstrates the vital importance of time-resolved search and follow-up of electromagnetic candidate counterparts of GW triggers. In such a scenario, the THESEUS mission, thanks to the diversity of intstrumentation, fast pointing and flexible schedule will represent an invaluable asset (see also Stratta et al. 2017).

*Acknowledgements.* The author acknowledge support from ASI grant I/004/11/3.

## References

Abbott, B. P., et al., 2017a, Physical Review Letters, 119, 161101

Abbott, B. P., et al., 2017b, ApJ, 848, L12

Alexander, K. D., et al., 2017, ApJ, 848, L21

Amati, L. et al., 2002, A&A 390, 81

Amati, L. et al., 2017, arXiv:1710.04638v2

Arcavi, I., Hosseinzadeh, G., Howell, D. A., et al. 2017, Nature, 551, 64

Berger, E., 2009, ApJ, 690, 231

Berger, E., 2011, NewAR, 55, 1

Berger et al. 2013, ApJ, 774, L23

Berger, E. 2014, ARA&A, 52, 43

Cantiello, M. et al. 2018, arXiv:1801.06080

Coulter, D. A., Foley, R. J., Kilpatrick, C. D., et al. 2017, Science, 358, 6370, 1556

Covino, S. et al. 2006, A&A, 447, L5

Cowperthwaite, P. S., et al., 2017, ApJ, 848, L17

D'Avanzo, P. et al. 2009, A&A, 498, 711

D'Avanzo, P. et al. 2014, MNRAS, 442, 2342

D'Avanzo, P. et al. 2015, JHEAp, 7, 73

D'Avanzo, P. et al. 2018, arXiv:1801.06164

Drout, M. R., Piro, A. L., Shappee, B. J., et al. 2017, Science, 358, 6370, 1570

Eichler, D., Livio, M., Piran, T., Schramm, D.N., 1989, Nature, 340, 126

Evans, P. A., Cenko, S. B., Kennea, J. A., et al. 2017, Science, 358, 6370, 1565

Fong, W. F. et al. 2013, ApJ, 769, 56

Fox, D. B., Frail, D. A., Price, P. A., et al. 2005, Nature, 437, 845

Ghirlanda, G., Nava, L., Ghisellini, G., Celotti, A., Firmani, C., 2009, A&A, 496, 585

Goldstein, A., et al., 2017, ApJ, 848, L14

Gottlieb, O., Nakar, E., Piran, T., & Hotokezaka, K. 2017, arXiv:1710.05896

Grindlay, J., Portegies Zwart, S. & McMillan, S., 2006, NatPh, 2, 116

Haggard, D., Nynka, M., Ruan, J. J., et al. 2017, ApJ, 848, L25

Hallinan, G., Corsi, A., Mooley, K. P., et al. 2017, Science, 358, 6370, 1579

Hjorth, J., Watson, D., Fynbo, J. P. U., et al., 2005a, ApJL, 630, L117

Hjorth, J., Watson, D., Fynbo, J. P. U., et al., 2005b, Nature, 437, 859

Hjorth, J., Bloom, J. S., 2012, Gamma-Ray Bursts, Cambridge Astrophysics Series 51, pp. 169-190

Hjorth, J., et al., 2017, ApJ, 848, L31

Jin et al. 2015, ApJ, 811, L22

Jin et al. 2016, Nat. Comm., 7, 12898

Kann, D. A., et al. 2011, ApJ, 734, 96

Kasliwal, M. M., Nakar, E., Singer, L. P., et al. 2017, Science, 358, 6370, 1559

Kathirgamaraju, A., Barniol Duran, R., & Giannios, D. 2017, arXiv:1708.07488

Kopac, D. et al., 2012, MNRAS, 424, 2392

Kouveliotou, C. et al., 1993, ApJ, 413, L101

Lazzati, D., Deich, A., Morsony, B. J., & Workman, J. C. 2017a, MNRAS, 471, 1652

Lazzati, D., Perna, R., Morsony, B. J., et al. 2017b, arXiv:1712.03237

Leibler, C. N., Berger, E. 2010, ApJ, 725, 1202

Li, L.-X., & Paczyński, B. 1998, ApJ 507, 59

Lipunov, V. M., Postnov, K. A., & Prokhorov, M. E. 2001, ARep, 45, 236

Lipunov, V. M., Gorbovskoy, E., Kornilov, V. G., et al. 2017, ApJ, 850, 1

Lyman, J. D., Lamb, G. P., Levan, A. J., et al. 2017, arXiv:1801.02669

Margutti, R. et al., 2013, ApJ, 756, 63

Margutti, R., et al. 2017, ApJ, 848, L20

Margutti, R., Alexander, K. D., Xie, X., et al. 2018, arXiv:1801.03531

Melandri, A., et al. 2017, GRB Coordinates Network, 21532

Mooley, K., Nakar, E., Hotokezaka, K., et al. 2017, arXiv:1711.11573

Nakar, E., 2007, Phys. Rev., 442, 166

Narayan, R., Paczynski, B., Piran, T., 1992, ApJ, 395, L83

Nicholl, M., et al., 2017, ApJ, 848, L18

Nicuesa Guelbenzu, A. et al. 2012, A&A, 548, A101

Pian, E., D'Avanzo, P., Benetti, S., et al. 2017, Nature, 551, 67

Rossi, E., Lazzati, D., & Rees, M. J. 2002, MNRAS, 332, 945

Ruan, J. J., et al., 2018, ApJ, 853, L4

Salafia, O. S., Ghisellini, G., Pescalli, A., Ghirlanda, G., & Nappo, F. 2015, MNRAS, 450, 3549

Salafia, O. S., Ghisellini, G., Ghirlanda, G., & Colpi, M. 2017, arXiv:1711.03112

Salafia, O. S., Ghisellini, G., and Ghirlanda, G., 2018, MNRASL, 474, 1, L7

Salvaterra, R. et al., 2008, MNRAS, 388, L6

Savchenko, V., et al., 2017, ApJ, 848, L15

Smartt, S. J., Chen, T.-W., Jerkstrand, A., et al. 2017, Nature, 551, 75

Soares-Santos, M., Holz, D. E., Annis, J., et al. 2017, ApJ, 848, L16

Stratta, G., 2017, arXiv:1712.08153

Tanvir et al. 2013, Nature, 500, 547

Tanvir, N. R., Levan, A. J., González-Fernández, C., et al. 2017, ApJ, 848, 27

Troja, E., et al. 2017b, Nature, 551, 71

Troja, E., et al. 2017a, GRB Coordinates Network, 22201

Troja, E., et al. 2018, arXiv:1801.06516

Valenti, S., David, J. S., Yang, S., et al. 2017, ApJ, 848, 24

Villar, V. A., Guillochon, J., Berger, E., et al. 2017, ApJ, 851, 21

Yang et al. 2015, Nat. Comm., 6, 7323

Yonetoku, D. et al., 2004, ApJ, 609, 935